    \documentclass[12pt]{article}
    \usepackage{psfig}  \usepackage{epsfig}  \usepackage{graphics}
    \usepackage{inputenc}
    \inputencoding{latin1}
     \newlength{\dinwidth}                       
     \newlength{\dinmargin}                      
\newcommand{\ZETF}[3]{{\it Zh.~Eksp.~Teor.~Fiz.} {\bf #1} ({#3}) {#2}}

\newcommand{\PLB}[3]{{\it Phys.~Lett.} {\bf B#1} ({#3}) {#2}}
\newcommand{\YF}[3]{{\it Yad.~Fiz.} {\bf #1} ({#3}) {#2}}
\newcommand{\SJNP}[3]{{\it Sov.~J.~Nucl.~Phys.} {\bf #1} ({#3}) {#2}}
\newcommand{\JETP}[3]{{\it Sov.~Phys.~JETP} {\bf #1} ({#3}) {#2}}
\newcommand{\NPB}[3]{{\it Nucl.~Phys.} {\bf B#1} ({#3}) {#2}}
\newcommand{\PRD}[3]{{\it Phys.~Rev.} {\bf D#1} ({#3}) {#2}}

\newcommand{\ZPC}[3]{{\it Z.~Phys.} {\bf  C#1} ({#3}) {#2}}

\newcommand{\CPC}[3]{{\it Comput.~Phys.~Comm.} {\bf #1} ({#3}) {#2}}
\newcommand{\JHEP}[3]{{\it J.~High~Energy~Phys.} {\bf #1} ({#3}) {#2}}

\def\lsim{\mathrel{\rlap{\lower4pt\hbox{\hskip1pt$\sim$}}
    \raise1pt\hbox{$<$}}}                
\def\gsim{\mathrel{\rlap{\lower4pt\hbox{\hskip1pt$\sim$}}
    \raise1pt\hbox{$>$}}}                
\def\ariadne{A\scalebox{0.8}{RIADNE}}
\def\jetset{J\scalebox{0.8}{ETSET}}
\def\lepto{L\scalebox{0.8}{EPTO}}
\def\herwig{\scalebox{0.8}{HERWIG}}
\def\ldcmc{\scalebox{0.8}{LDCMC}}
\def\rapgap{R\scalebox{0.8}{AP}G\scalebox{0.8}{AP}}
\def\smallx{\scalebox{0.8}{SMALLX}}
\def\smmod{\scalebox{0.8}{SMMOD}}
\def\cascade{C\scalebox{0.8}{ASCADE}}
\newcommand{\ee}{\mbox{${\rm e^+e^-}$}}
\begin{document}
\begin{flushright}
  GLAS-PPE/1999-10\\
  LU TP 99--21\\
  hep-ph/9908367 \\
  August 1999
\end{flushright}
\vspace*{10mm}
\begin{center}  \begin{Large} \begin{bf}
      Summary of activities in the Working Group on\\
      QCD Cascades of the HERA Monte Carlo workshop\\
    \end{bf}  \end{Large}
  \vspace*{5mm}
   {\it To be published in the proceedings of the\\
   HERA Monte Carlo workshop, Hamburg, Germany, 1998-1999}

   \vspace*{5mm}
  \begin{large}
    Nick Brook$^a$, Leif L\"onnblad$^b$\\
  \end{large}
\end{center}
$^a$ Dept.\ of Physics \& Astronomy, University of Glasgow, Glasgow, UK.\\ 
$^b$ Department of Theoretical Physics, Lund University, Lund, Sweden\\
\begin{quotation}
  \noindent
  {\bf Abstract:} We summarize the activities in working group 10
  concerned with QCD cascades, and find that although much work still
  needs to be done, much progress was made during this workshop
  in understanding the merits and deficiencies of different programs.
\end{quotation}

\section{Introduction}

The importance of theoretically well founded event generators which
describe experimental data to a satisfactory degree can in general not
be overstated. But since the whole of this workshop is dedicated to
Monte Carlo event generators, we feel it is unnecessary to dwell on
this here.

Event generators based on perturbative QCD cascade models have been
immensely successful in reproducing the bulk of the data recorded at
LEP \cite{LEPMC}. Some of us were hoping for a similar development at
HERA, but it soon became clear that this was not possible.

Before this workshop there was not one QCD cascade based generator
which even came close to the agreement with data achieved at LEP.
This was especially true for deep inelastic scattering (DIS) at small
$x$: none of the conventional generators with DGLAP-based \cite{DGLAP}
initial state cascades was able to even qualitatively describe the
measured final state in the proton direction \cite{hzsmallx}. It was
the hope that generators based on BFKL \cite{BFKL} or CCFM \cite{CCFM}
evolution could be developed and would be able to describe this
region, but the only two such generators available before the
workshop, \smallx\ \cite{smallx} and \ldcmc\ \cite{LDCMC}, did not
live up to these expectations.

The only generators which gave a fair description of small-$x$ final
states were generally not considered to be on firm theoretical
grounds: \rapgap\ modeling a resolved virtual photon contribution
\cite{RAPvirt}, and \ariadne\ \cite{ariadne} based on the colour
dipole cascade model \cite{cdm}.

One could hope that the generators at least would give a good
description of data at high $Q^2$ and high $x$, where
DGLAP evolution should be a good approximation of the underlying
parton dynamics. But most of the generators were not even able to
reproduce data in this region \cite{hzhiq2}.

With this in mind the work of our group was quickly divided into two
directions. One was concerned with developing new generators
implementing CCFM evolution to better understand the small $x$ region.
The other direction was to look at existing generators and try to
understand why some of them failed to describe the high $Q^2$ region
and, if these discrepancies could be fixed, try to tune the parameters
in the programs to get as good description as possible of available
data.

\section{Confronting the Generators}

\subsection*{High $Q^2$ and the Current Breit Hemisphere}

Over the course of the workshop much understanding and development of
the event generators and the data in the high $Q^2$ region and the current
fragmentation region of the Breit frame has been achieved.

One of the puzzles before the workshop was why the \ariadne\ program,
which gives a good overall description of the HERA DIS data, had
problems describing the high $Q^2$ region. In this region one na\"{\i}vely
expected our understanding of the underlying physics to be on a firmer
theoretical base than at low $x$ and $Q^2.$ Even in the current
fragmentation region of the Breit frame, where it is expected that
DIS events should resemble one hemisphere of \ee\ annihilation events,
\ariadne\ had difficulties in describing the data. This is despite the
fact that \ariadne\ gives a very good description of \ee\ data.

The main difference between the treatment of colour dipoles in \ariadne\
between \ee\ and DIS is due to the fact that in DIS the initial parton
configuration is not point like; the proton remnant is treated as an
extended object. It was observed~\cite{lonnblad} at high $Q^2$ the
phase space available for radiation was restricted even in the current
fragmentation region due to this treatment 
of the extended proton remnant. This deficiency in the model has been
understood and overcome~\cite{lonnblad} and modifications introduced
within \ariadne. These modifications went a long way towards removing
the discrepancies between \ariadne\ and the data, though problems
still persist in describing the data and are under 
investigation~\cite{eduardo}.

\emph{Ed\'en} questioned the assumptions behind the equivalence of the
current fragmentation region of the Breit frame and a single
hemisphere in \ee\ experiments~\cite{eden}.  It was shown that in DIS
QCD radiation can give rise to high $p_T$ emissions which have no
correspondence in an \ee\ event; these emissions lead to a
de-population of the current fragmentation region. In order to limit
the effect of these high $p_T$ emissions, a jet algorithm was applied
and DIS events with a jet $p_T > Q/2$ were removed from the
comparison. This had a sizeable reduction in the discrepancy between
the predictions of the mean charged multiplicity of \ee\ and DIS Monte
Carlo generators at low $Q^2$. This cut on jet $p_T$ though suppresses
the contribution to the DIS sample from boson-gluon fusion. In so
doing, charm production is reduced by approximately a factor of 2.
Further improvement between the \ee\ and DIS generators could then be
achieved by artificially removing heavy quark contributions from the
\ee\ generator by generating events with just light quarks. It was
proposed that the experiments perform further studies of the current
fragmentation region, applying this jet selection criteria, to compare
with light quark enriched samples from the LEP1 experiments.

A comparison between MLLA predictions and the ARIADNE Monte Carlo at
the parton level were
made during the workshop~\cite{brook}. 
It was shown for the \ee\ scenario there was a
good agreement between the MLLA truncated parton spectra and that
generated from \ariadne. The MLLA predictions for the current region of
the Breit frame in DIS are identical to the \ee\ predictions, so the
study was extended to DIS. It was shown that a reasonable agreement only
became possible with the introduction of the previously discussed
modifications of the proton remnant in \ariadne.
At lower $Q^2$ discrepancies between the
MLLA predictions and \ariadne\ were observed. These discrepancies could
well be due to the problems discussed by \emph{Ed\'en}~\cite{eden}. 

\subsection*{Comparisons with Data}

During the workshop a forum was established between the H1 and ZEUS
collaborations for a joint coordinated investigation of the generators,
working closely with the programs' authors. 
The HzTool package~\cite{hztool} was substantially updated
to include the majority of H1 and ZEUS hadronic
final state data both for DIS and photoproduction. A concerted effort
was also made to include preliminary data from the collaborations in
this program.

During the course of the workshop there were two investigations of the
Monte Carlo predictions compared to the HERA DIS data~\cite{eduardo,guenter}.
One attempted to `tune' the program's in the higher $(x,Q^2)$
region~\cite{eduardo}. In this study
the \ariadne~\cite{ariadne}, \herwig~\cite{herwig} and \lepto~\cite{lepto}
Monte Carlo generators for DIS data were
investigated~\cite{eduardo}.
Other programs such as \rapgap~\cite{rapgap} and
those developed over the duration of the workshop are planned to
be examined as part of a continuing program of work by the authors.
The other study~\cite{guenter}
concentrated on a comparison with new energy flow~\cite{eflows} data
and particle spectra data and, in addition, included a
comparison of the data with the \rapgap\ program.

Both studies found that modifications to the generators (e.g.\ new
soft colour interaction implementation in \lepto\ and the high $Q^2$
changes in \ariadne) helped tremendously in trying to describe the
data.  Unfortunately, it proved difficult to find sets of parameters
within the generators studied that would describe the whole range of
distributions at both low and high $Q^2.$ As a compromise various
setting have been given, optimised for particular regions of phase
space.

The momentum generated by the workshop in trying to `tune' the
generators will allow more detailed investigations, including generators
not so far studied. The ultimate aim of the forum, set up as a
consequence of the
workshop, is to have event generators that are able to describe the
HERA data  as impressively as they do the LEP data !

\section{Developing New Generators for small $x$}

The fact that DGLAP-based generators have not been able to reproduce
small-$x$ DIS final states measured at HERA has often been taken as an
indication that effects of BFKL or CCFM evolution is visible in the
data. This view has been strengthened since the \ariadne\ generator,
which has the feature of $k_\perp$-non-ordering in common with BFKL,
qualitatively describes the data.

To really verify that BFKL evolution is responsible for e.g.\ the high
rate of forward jets, or the large forward transverse energy flow, it
is necessary to have an event generator with BFKL dynamics properly
implemented. One such generator, \smallx\ \cite{smallx}, has been
available for quite some time. It implements CCFM evolution, which in
the small-$x$ limit is equivalent to BFKL, but it could only generate
events at the parton level.

Not long before the workshop, the \ldcmc\ generator was released which
implements the Linked Dipole Chain model, a reformulation of the CCFM
evolution. Although this generator describes the small-$x$ region
slightly better than DGLAP-based ones, it was clearly not able to
explain the data in the forward region to a satisfactory level
\cite{LDCMC}.

Before the workshop it was already clear that small-$x$ DIS final
states could be described by adding to a normal DGLAP-based generator,
a contribution corresponding to a resolved virtual photon. This model
was treated in detail in Working Group 30, while our group
concentrated on developing CCFM-based models. So, although resolved
virtual photons may be a reasonable way to describe small-$x$ final
states we will not discuss them further here.

During this workshop, a lot of effort was put into developing old and
new generators based on CCFM evolution. \emph{Goerlich and Turnau}
have developed the \smallx\ generator so that it is now interfaced to
the Lund string fragmentation model implemented in \jetset\ 
\cite{jetset}. Using a simple parameterization of the input gluon
density they can evolve $F_2$, but fail to find a good description of the
HERA data.  Regardless of this poor agreement, they use this gluon
parameterisation to generate the hadronic final state. They find that
they cannot describe the transverse energy flow, but contrary to
\ldcmc\ they overshoot the data rather than undershoot it
\cite{lidia}.

Much effort has also been put into trying to understand the
discrepancies between the \ldcmc\ and data. The LDC model should, to
leading order, be equivalent to CCFM, but the program also includes
estimates of non-leading effects and has e.g.\ included the evolution
of quark chains and the correction of splitting functions to reproduce
$2\rightarrow2$ matrix elements for local hard sub-collisions in the
ladder. However, no significant progress was made during this
workshop, and \ldcmc\ still does not reproduce data at small $x$.

Similarly disappointing results were presented by \emph{Salam}
\cite{salam}, who reported on the work done by the Milan group on CCFM
phenomenology.  They investigated different possible formulations of
the so-called non-Sudakov form factor \cite{CCFM}. Although their
modifications were formally sub-leading and only important as
$z\rightarrow1$ in the splittings, large effects were noticed. With
the modification which led to the largest correction, they were able
to reproduce the $F_2$ measured at HERA, but not the final state
properties, such as forward jet rates. It should be noted that they
did not try to include the, formally sub-leading, $1/(1-z)$ pole in
the splitting function, which also could give rise to large
corrections.

Also \emph{Jung} \cite{hannes} has used a somewhat non-standard form
of the non-Sudakov form factor introduced in \cite{kincon}.
Implementing this in the \smallx\ program\footnote{Together with some
  other modifications, the obtained version of \smallx\ is called
  \smmod.} he obtains a good description of $F_2$. For the final state
properties, he finds a large dependency on the so-called
\textit{kinematical}, or \textit{consistency constraint}, which was
introduced ensure that the standard form of the non-Sudakov form
factor is below unity in the allowed phase space. Since the
non-Sudakov used by \emph{Jung} does not suffer from this problem, the
consistency constraint is not needed, and without it a good
description of the data, e.g.\ forward jet rates, is obtained.

From the \smallx\ program it is possible to extract the evolved
unintegrated gluon density. This is used by \emph{Jung} in a
completely new program, \cascade, which implements CCFM in a backward
evolution framework \cite{hannes}. Also with this program a reasonable
description of small $x$ data is obtained, although the agreement
between \cascade\ and \smmod\ is not perfect due to differences
in the normalizations of the unintegrated gluon density between
the two programs.

Clearly much progress has been made during this workshop, although
much work is still needed. The fact that there exist three hadron-level
generators which all claim to (in leading order) implement the same
CCFM evolution, but giving completely different results, urgently
calls for further investigations. In comparing these models among
themselves and with data, it is important to have good observables
which are sensitive to the characteristics of BFKL/CCFM evolution. Two
new such observables was suggested during the workshop.
\emph{Goerlich and Turnau} suggested measuring the difference in
transverse energy flow with and without a forward high-$k_\perp$
particle. This showed a good separation power between the DGLAP-based
\lepto\ generator, their own \smallx\ generator and \ariadne\ 
\cite{lidia}. A similarly good separation power was shown for some
observables based on transverse momentum transfer proposed by
\emph{Van Mechelen and De Wolf}. By summing vectorially all transverse
momentum on either side of a given rapidity cut, the $k_\perp$ of the
propagator gluon is reconstructed, and correlations can be measured as
function of rapidity. In this way it should be possible to test the
$k_\perp$-non-ordering property of BFKL evolution \cite{pierre}.

More general advice to event generator authors, was given by
\emph{Levin} \cite{levin}, who discussed the recently calculated
next-to-leading corrections to BFKL \cite{NLLBFKL}, and their
implications for small-$x$ evolution. He also presented indications
that so-called screening corrections due to the large gluon density at
small $x$, are becoming visible at HERA, and urged Monte Carlo authors
to consider including such corrections in their programs.

\section{Conclusions}

As far as QCD cascades are concerned, this has been a very productive
workshop. During the workshop we have increased our understanding of
the high $Q^2$ region and why the standard generators had problems
describing the corresponding hadronic final states. Now, most of these
problems have been solved and attempts to tune the parameters of these
generators has started. But many difficulties still remains, and it
has not been possible to find a single consistent set of parameters
for any of the generators which can describe all observables at high
$Q^2$.

At small $x$ the situation is even worse. But also here much work has
been invested and much progress has been made during the workshop.
There are now three hadron-level generators implementing CCFM
evolution. Although only one of them has been shown to be able to
reproduce the characteristics of small-$x$ final states, the situation
is much better than before the workshop, when only the \ariadne\ and
\rapgap\ (with resolved virtual photons) programs gave a reasonable
description. By carefully comparing all these different models we may
soon be able to understand better the underlying parton dynamics.

Although the workshop is now over, the work will continue and so will
the fruitful collaboration between experimentalists and event
generator authors.

\end{document}